# Analysis of Phase Formations and Mechanical Properties in Complex Concentrated Alloys by Machine Learning Approach

Jie XIONG[b,c], San-Qiang SHI[b,c]*, Tong-Yi ZHANG[a]*

[a]Material Genome Institute, Shanghai University, Shanghai, China

[b]Department of Mechanical Engineering, The Hong Kong Polytechnic University, Hong Kong, China

[c]Shenzhen Research Institute, The Hong Kong Polytechnic University, Shenzhen, China

**Abstract**

The mechanical properties of complex concentrated alloys (CCAs) depend on their forming phases and corresponding structures, the prediction of the phase formation for a given CCA is essential to its discovery and applications. 541 sample were collected from previous studies, comprising 61 amorphous, 164 single-phase crystalline, and 361 multi-phases crystalline CCAs. We proposed three classification models to category and understand the phase selection of CCAS. Also, a two-objective regression model was constructed to predict the hardness and compressive yield stress of CCAs. All three classification models have accuracies higher than 85%, and correlation coefficient of random forest regression model is greater than 0.9 for both of two objectives. In addition, we proposed four descriptors via multi-task SISSO method to predict the mechanical properties of CCAs, the average correlation coefficient of SISSO models is higher than 0.85. The present work demonstrates the great potential of machine learning approach in the prediction of target properties in CCAs.

## 1. Introduction

The conventional metal alloys contain only one principal element, their mechanical properties are generally dominated by the principal elements[1]. A new branch of metal alloys, complex concentrated alloys (CCAs) with excellent mechanical properties and unexpected microstructures have been developed in the past decades prior to 2004. CCAs encompass medium-entropy alloys (MEAs) consisting of three or four principal components and high-entropy alloys (HEAs) consisting of more than four principal components[2–7].

According to the Gibbs phase rule $F = C - P +1$ (where $F$ is the degree of freedom, $C$ is the number of components and $P$ is the number of phases), multiple elements might produce many ($C + 1$) equilibrium phases in CCAs, a maximum of six equilibrium phases are expected in the case of a five-component CCA[8]. However, the high configuration entropy in a CCA can enhance the formation of a single phase rather than multiple phases[3,9]. Many CCAs were reported to only form single-phase disordered solid solutions on simple body-centered-cubic (BCC)[10], face-centered-cubic (FCC)[11] and hexagonal close-packed (HCP)[12] lattices. In addition, some CCAs were found to form single intermetallic phase, such as C14 Laves phase[13], and B2 phase[14].

These simple structures limited the applications of CCAs, structural and functional materials thus require multiple phases to achieve balanced properties. There have been multi-phase CCAs that only consist of disordered solid-solutions[15], and multi-phase CCAs that contain intermetallic phases[16]. Besides, some amorphous CCAs have been developed in the past decades[17]. A new CCA thus might adopt among three possible situations: amorphous (AM), single-phase (SP), and multi-phases (MP).

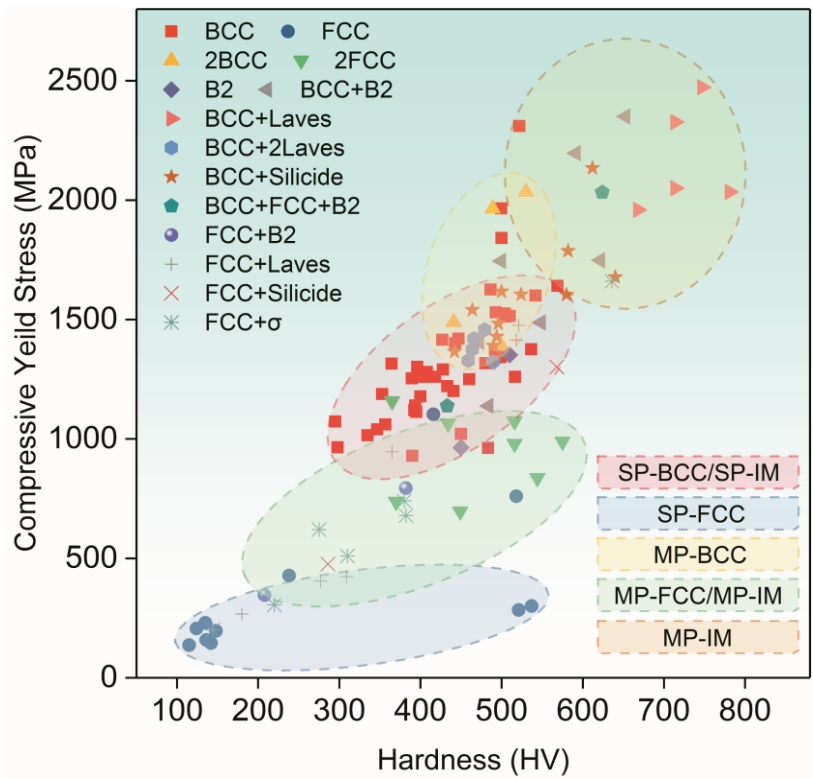

Figure 1. Ashby plot of compressive versus hardness of complex concentrated alloys (CCAs) with different resulting phases. The mechanical properties of CCAs highly depend on their phases.

Figure 1 shows that the mechanical properties of CCAs depend on their forming phases,

the MP CCAs are stronger than the SP CCAs, the CCAs that contain intermetallic phases (termed as IM) are stronger than the CCAs that only consists of solid solutions (termed as SS), the SP-BCC and MP-BCC CCAs are harder and stronger than the SP-FCC and MP-FCC CCAs.

Hence, the prediction of the phase formation for a given CCA is essential to the discovery and applications of new alloys. It is agreed that the phase formation correlates with parameters such as the mismatch in atomic size ($\delta_{MR}$) and in the Pauling electronegativity difference ($\delta_{XP}$), the valence electron concentration ($\overline{VEC}$), the entropy of mixing ($S_{mix}$), the enthalpy and mixing ($H_{mix}$), and the omega parameter ($\Omega$)[18–20]. These correlations were given via summarizing a small set of experimental data, which might limit the generalization ability of such correlations to a broader field of CCAs.

Thermodynamic modeling, density functional theory (DFT), and molecular dynamics (MD) are useful techniques to simulate CCAs. Gorsse et al.[21] applied the CALPHAD method to predict the phase formation of CCAs, Huhn et al.[22] utilized the DFT approach to predict the phase transformation in CCAs. However, the number of elements in CCAs gives rise to the complex and computational cost of those simulating techniques.

Recently, the materials community has utilized machine learning tools to deep-research materials, including steels[23–25], metallic glasses[26–30], shape memory alloys[31], and CCAs[32–34]. Wen et al.[32] proposed a ML-based strategy to find new CCAs with high hardness in the Al-Co-Cr-Cu-Fe-Ni system. For phase selection, Islam et al.[33] employed a neural network to classify the corresponding phase selection in 118 CCAs with an average test accuracy of higher than 80%. Huang et al.[34] adopted and compared three different ML algorithms on 401 CCAs to predict the phases. The prediction of yield stress was absent in these studies, the formation of disordered solid solutions in CCAs was not mentioned either.

In this work, we proposed a ML framework to investigate the phase formation ability and predict the mechanical properties of CCAs. A dataset of CCAs was created by collecting data from related studies. These data covered variables such as processing conditions, resulting phases, hardness and compressive yield stress. Different features were introduced, and various feature selection algorithms were utilized to select key features during model construction. The random forest (RF) algorithm and principal component analysis (PCA) was used to demonstrate the predictive power of selected key features. Besides, we used random forest regression algorithm to predict the hardness and compressive yield stress of CCAs.

## 2. Methodology

### 2.1 Data Collection and ML Algorithm

The forming phases of 1042 CCAs under various conditions such as as-cast, as-annealing, and mechanical-alloying are collected from several related studies. The previous studies show that the phases formed in the as-cast state are more stable and similar to the equilibrium state[35]. Thus, the datasets are assembled from 750 as-cast CCAs fabricated by vacuum arc melting only. The individual samples with conflict reports on phases are excluded. For example, the alloy $Al_{0.3}CoCrFeMo_{0.1}Ni$ is not considered because it is reported as SP-FCC and FCC+L1$_2$ in two different studies, the alloy $Al_{0.8}CrCuFeNi_2$ is not considered because it is reported as SP-FCC and BCC+FCC in two different studies. Finally, the training dataset in the Supporting Information consists of 541 unique alloys compositions in total, comprising 61 AM samples, 133 SP-SS samples, 31 SP-IM samples, 92 MP-SS samples and 224 MP-IM samples. In addition, the average value of conflict reports on mechanical properties are adopted,

An efficient and robust ML algorithm available in the WEKA library[36], random forest (RF), is employed for model construction. All built models are evaluated via a ten-fold cross-validation method.

### 2.2 Feature Generation and Selection

The collected data had to be transformed into features that can be recognized as the input of ML algorithms. A set of features should be represented to ensure that a well-performing ML model was generated. Here, the manual features (MFs) were employed for model construction. The MFs of an alloy were complied with the elemental statistics, valence electron distributions, and thermodynamic features. The elemental statistics were calculated from the elemental properties shown in Table 1 based on the following equations.

$$\bar{x} = \sum a_i x_i \tag{1}$$

$$\delta_x = \sqrt{\sum a_i (1 - x_i/\bar{x})^2} \tag{2}$$

where $a_i$ and $x_i$ is the atomic fraction and elemental properties of the $i$ constituent, respectively. The value of $\delta_x$ was set to zero if the value of $\bar{x}$ was zero.

The valence electron distributions comprised the fraction ($f_{s,p,d}$) of the electrons in the *s*, *p*, *d* valence orbitals of an element, as given by equation

$$f_{s,p,d} = \overline{(s,p,d)\text{VEC}} / \overline{\text{VEC}} \tag{3}$$

Table 1. The used 18 basic elemental properties and their values can be seen in Table S2

| Elemental Property (Abbreviation) | | |
| --- | --- | --- |
| Atomic Number (*AN*) | Melting Point (*Tm*) | Mulliken Electronegativity (*XM*) |
| Metallic Radius (*MR*) | Boiling Point (*Tb*) | Pauling Electronegativity (*XP*) |
| Valence Electrons (*VEC*) | Heat Capacity (*C<sub>p</sub>*) | First Ionization Potential (*I*1) |
| VEC in the *s* orbital (*s*VEC) | Thermal Conductivity (*K*) | Second Ionization Potential (*I*2) |
| VEC in the *p* orbital (*p*VEC) | Heat of Latent (*H<sub>L</sub>*) | Electron Affinity (*Eea*) |
| VEC in the *d* orbital (*d*VEC) | Lattice Volume (*LP*) | Work Function (*W*) |

Thermodynamic features include the enthalpy of mixing ($H_{\text{mix}}$) based on Miedema's empirical method, the enthalpy of formation ($H_f$) based on DFT calculations given in the open quantum materials database, the entropy of mixing ($S_{\text{mix}}$), the entropy of latent ($S_L$), the Gibbs free energy of mixing ($G_{\text{mix}}$), two self-defined parameter *Ω'* and *α*

$$H_{\text{mix}} = 4 \sum_{i=1}^{N} \sum_{j=1}^{N} \Delta H_{ij} a_i a_j \tag{4}$$

$$S_{\text{mix}} = -R \sum_{i=1}^{N} a_i \ln a_i \tag{5}$$

$$S_L = \overline{H_L} / \overline{T_m} \tag{6}$$

$$G_{mix} = H_{\text{mix}} - \overline{T_m} S_{\text{mix}} \tag{7}$$

$$\Omega' = |H_{\text{mix}}| / \overline{T_m} S_{\text{mix}} \tag{8}$$

$$\alpha = H_f - H_{\text{mix}} \tag{9}$$

In summary, 45 manual features were generated and normalized to [0, 1] for further selection. Redundant features are removed based on the Pearson correlation coefficient (PCC) and information gain algorithm (IG), values of PCC bigger than 0.8 or smaller than -0.8 indicate a strong linear correlation between two different features, the one gains less information are removed.

Then four different feature selection methods, including the sequential backward selection

(SBS), sequential forward selection (SFS), IG and RF are employed and compared to screen these nonlinear-correlated features. In general, a good ML model should provide a low miss rate with as less as a possible number of features to achieve a balance between accuracy and complexity. Two objective functions, the miss rate ($OF_1$) and the number of features ($OF_2$) (both are scaled to [0.2, 0.8]) should be optimized. The subset of features which minimize the product of $OF_1$ and $OF_2$ will be the Pareto optimal result.

## 3. Results and Discussion

The five-label classification (AM, SP-SS, SP-IM, MP-SS, and MP-IM), whose miss rate exceeds 20% is not suitable. We therefore apply ML algorithms to two classifications, a ternary classification (AM, SP, and MP) termed as ML-A, and a binary classification (SS and IM for crystalline CCAs) termed as ML-B. Besides, we proposed a ternary classification termed as ML-C to predict the structure of solid solutions as SP or MP BCC (all labeled as BCC), SP or MP FCC (labeled as FCC), and the mixture of BCC and FCC (labeled as MSS), the CCAs contain HCP are excluded because of only 5 related samples.

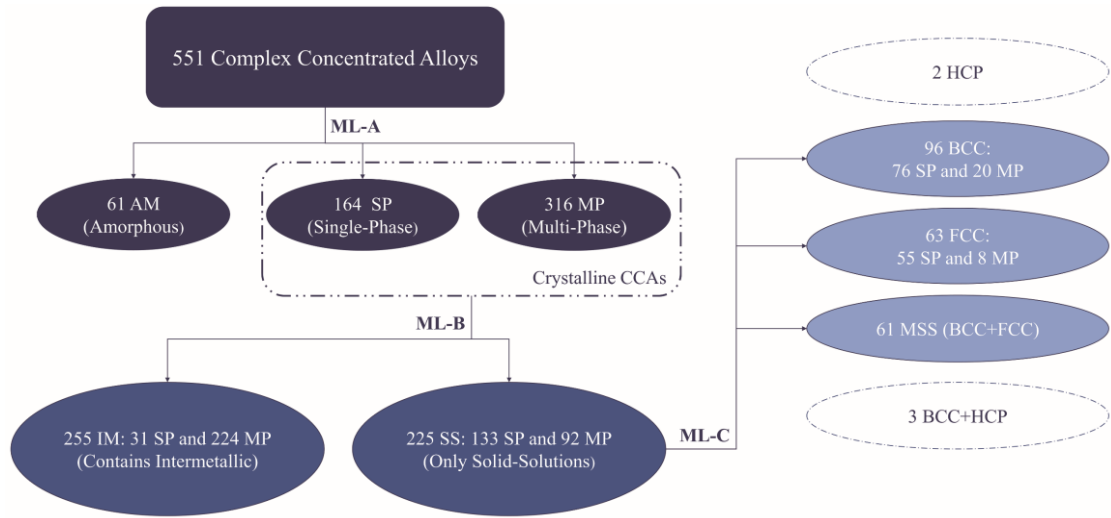

Figure 2. Three classifications in this work, a ternary classification termed as ML-A to classify AM, SP and MP, a binary classification termed as ML-B to classify SS and IM, and a ternary classification termed as ML-C to classify BCC, FCC, and MSS.

### 3.1 Optimal Feature Subset

First, fifteen, nineteen and twenty-two redundant features are removed by PCC shown in

Supporting information as Figures S1-S3 for ML-A, ML-B, and ML-C classification, respectively. The remaining features are listed in the Supporting Information as Table S2.

Then, for ML-A classification, Figure 3(a) shows the miss rate of each RF model as a function of the number of features, each point stands for an RF model trained with a subset containing a certain number of features selected by different feature selection method. It can be observed that the RF model with a subset containing 13 features has the lowest miss rate as 12.38%, while the RF model with a subset containing six features has a miss rate of 14.05% and minimize the product of two optimal functions shown in Figure 3(b). The optimal subset is termed as OS-A, which contains the average boiling temperature ($\overline{Tb}$), the average valence electron concentration in the $d$ orbital ($\overline{dVEC}$), the mismatches in metallic radius ($\delta_{MR}$) and in boiling temperature ($\delta_{Tb}$), the enthalpy of formation ($H_f$) and the entropy of latent ($S_L$).

For ML-B classification, the RF model with a subset containing 14 features is found to have the lowest miss rate as 10.83% in Figure 3(c), while the optimal subset in Figure 3(d) only contains five features and the corresponding RF model has a miss rate of 12.71%. The optimal subset OS-B contains the average boiling temperature ($\overline{Tb}$), the average second ionization potential ($\overline{I2}$), the mismatches in the first ionization potential ($\delta_{I1}$) and in the heat capacity ($\delta_{Cp}$), and the enthalpy of formation ($H_f$).

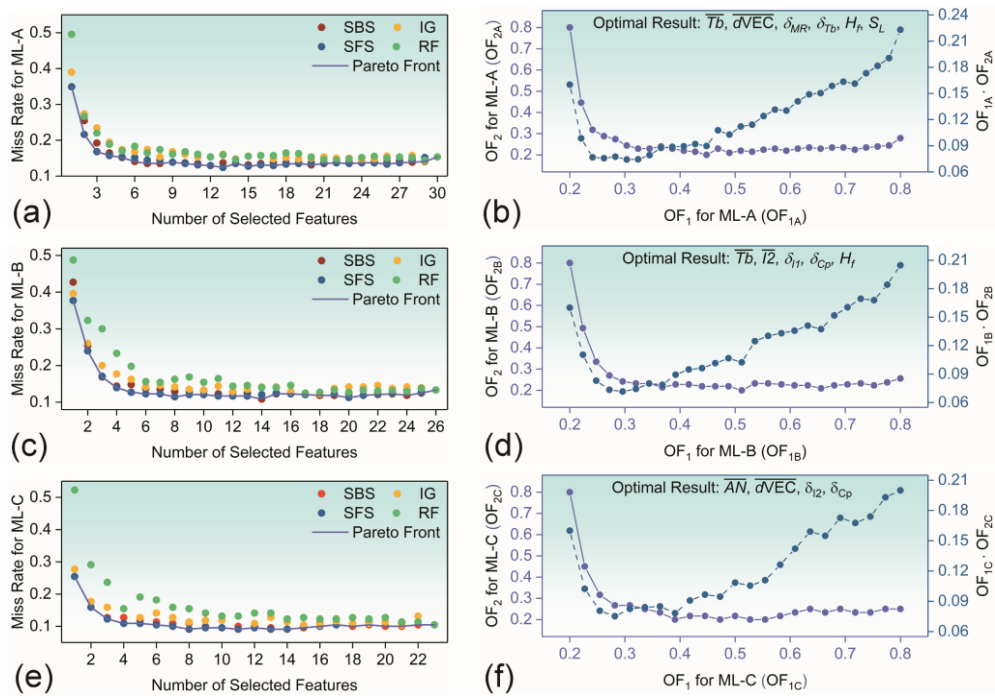

Figure 3. The process of feature selection and random forest (RF) model construction for phases

classification in CCAs. (a) plots the miss rate of random forest model with 100 trees for ML-A classification against a certain number of features selected by the sequential backward selection (SBS), the sequential forward selection (SFS), the information gain (IG), and RF method. (b) indicates that the optimal subset contains six features ($\overline{Tb}$, $\overline{dVEC}$, $\delta_{MR}$, $\delta_{Tb}$, $H_f$, $S_L$) achieves the balance of model accuracy and complexity for ML-A classification. (c) plots the miss rate of random forest model with 100 trees for ML-B classification against a certain number of features selected by SBS, SFS, IG, and RF method. (d) indicates that the optimal subset contains five features ($\overline{Tb}$, $\overline{I2}$, $\delta_{I1}$, $\delta_{Cp}$, $H_f$) achieves the balance of model accuracy and complexity for ML-B classification. (d) plots the miss rate of random forest model with 100 trees for ML-C classification against a certain number of features selected by SBS, SFS, IG, and RF method. (d) indicates that the optimal subset contains four features ($\overline{AN}$, $\overline{dVEC}$, $\delta_{I2}$, $\delta_{Cp}$) achieves the balance of model accuracy and complexity for ML-C classification.

For ML-C classification in Figures 3(e) and 3(f), the lowest miss rate of 9.09% appears when the RF model built with a subset containing 8 features, the subset containing four features is the optimal result OS-C, with which the RF model has a miss rate of 10.91%. OS-C contains the average atomic number ($\overline{AN}$), , the average valence electron concentration in the $d$ orbital ($\overline{dVEC}$), the mismatches in the second ionization potential ($\delta_{I2}$) and in the heat capacity ($\delta_{Cp}$).

Table 2. Features in the optimal subset for ML-A, ML-B, and ML-C classification.

| Classification | Classes | Features in the optimal subset |
| --- | --- | --- |
| ML-A | AM, SP, MP | $\overline{Tb}$, $\overline{dVEC}$, $\delta_{MR}$, $\delta_{Tb}$, $H_f$, $S_L$ |
| ML-B | SS, IM | $\overline{Tb}$, $\overline{I2}$, $\delta_{I1}$, $\delta_{Cp}$, $H_f$ |
| ML-C | BCC, FCC, MSS | $\overline{AN}$, $\overline{dVEC}$, $\delta_{I2}$, $\delta_{Cp}$ |

## 3.2 Parameters and performance of random forest

Random forest is a parallel ensemble learning approach using decision tree as the base learner. We build random forest with 20 trees for ML-A classification, the ML model is termed as RF-A and 468 samples are correctly classified by the RF-A model. The receiver operating characteristic (ROC) curves and corresponding aera under the curve (AUC) of RF-A model are shown in Figure 4(a), the micro-average AUC value of RF-A model is 0.91. The forest with 160 trees termed as RF-B is grown for ML-B classification, 195 SS and 225 IM samples are correctly classified, and the micro-average AUC value can reach as high as 0.931. As for ML-C classification, the 260 decision trees are used for composing random forest, structures of 196 samples are correctly predicted, the micro-average AUC value of built forest can reach 0.973.

AUC value usually varies between 0.5 and 1, a value bigger than 0.9 represents an outstanding classifier, the great values of AUC indicate that three built RF models perform excellently.

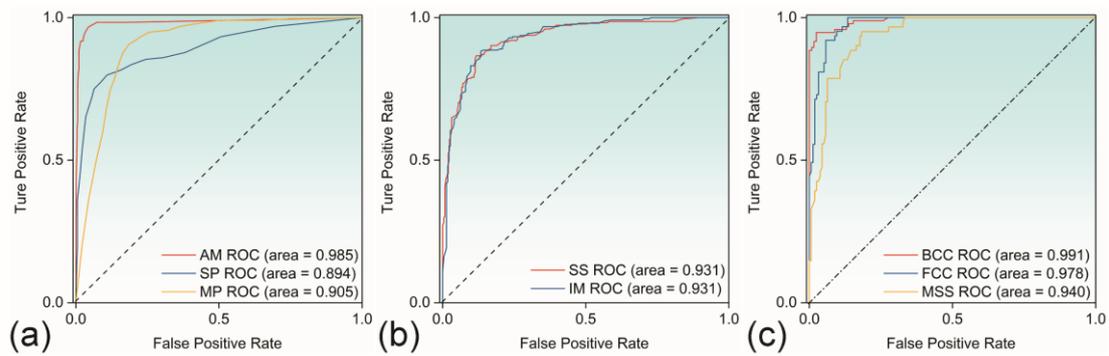

Figure 4. The receiver operating characteristic (ROC) curves of random forest for different classifications with the area under the ROC curves. (a) shows the ROC curves of random forest with 20 trees for ML-A, (b) ROC curves of random forest with 160 trees for ML-B, (c) shows the ROC curves of random forest with 260 trees for ML-C.

## 3.3 Visualization of Selected Features

Visualization is a good way to show the predictive power of our selected features. However, visualizing high dimensional features is very tough. Features selected before are thus compressed by principal component analysis (PCA), and the first two principal components of them are visualized in Figure 5. The AM samples, SP samples, and MP samples cluster in the red, blue, and yellow regions of Figure 5(a), respectively. These two principal components can recognize those samples well. Four features in the OS-B subset are also transformed into principal components by PCA, SS samples in red and IM sample in bule are instinctively separated in the transformed space shown in Figure 5(b). Besides, three features in the OS-C subset are compressed to PCs as shown in Figure 5(c), the BCC structures can be distinguished well from other structures by the first PC, and the second PC can category FCC structures and

MSS. The PCA results indicate that those features selected before have strong predictive power on different classifications.

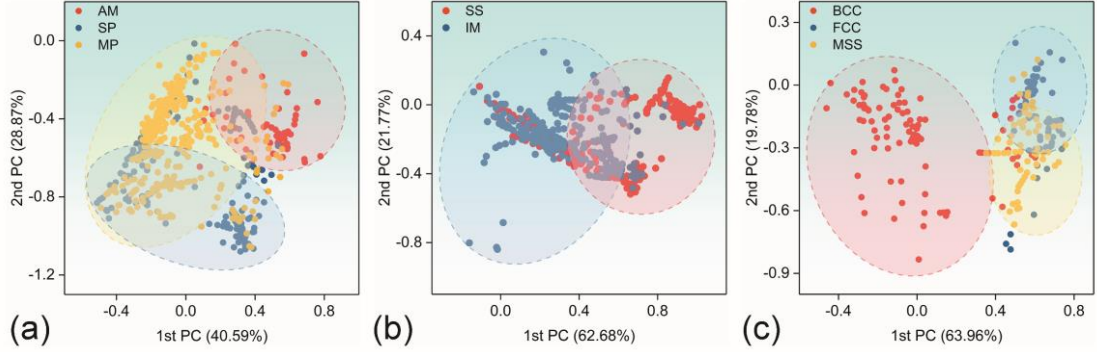

Figure 5. The first two principal components generated with features listed in Table 2 for (a) 541 CCAs with AM in red, SP in bule and SP in yellow; (b) 480 crystalline CCAs with SS in red and IM in bule; (c) 220 CCAs with structures as BCC in red, FCC in blue and MSS in yellow. The percentage is the ratio of the variance on the corresponding principal axis direction.

### 3.4 Prediction of mechanical properties

As mentioned before, the mechanical properties such as hardness and compressive yield stress of CCAs highly depends on their forming phases and structures. Six, five, and four features listed in Table 2 were selected for ML-A, ML-B, and ML-C classification, respectively. Thus, we got eleven individual features in total which affect the phase formations of CCAs, and these eleven features are probably used for predicting the mechanical properties of CCAs as well.

We employ multi-objective random forest regression (MO-RFR) to build ML models to predict the relative hardness ($RH$) and relative compressive yield stress ($RYS$) shown as following equations:

$$RH = H/\overline{AW} \tag{10}$$

$$RYS = YS/\overline{AW} \tag{11}$$

in which $H$ and $YS$ is, respectively, the measured hardness and compressive yield stress, $\overline{AW}$ is the average atomic weight.

The average correlation coefficient is adopted here to select features that can affect both of RH and RYS, that is, the subset with highest average correlation coefficient ($R$) is selected at each step of SBS and SFS procedure. It can be seen in the Figures 6(a) that the average correlation coefficient shows little improvement beyond five features and become even worse when exceed seven features. Hence, the subset for predicting the RH and RYS contains five features selected by SBS, namely the average atomic number ($\overline{AN}$), the average second ionization potential ($\overline{I2}$), the average valence electron concentration in the $d$ orbital ($\overline{dVEC}$), the mismatches in the second ionization potential ($\delta_{I2}$) and in the metallic radius ($\delta_{MR}$). Figures 6(c) and 6(d) show, respectively, the predicted RH and RYS against the actual value, indicating well agreement, as evidenced by $R > 0.9$ for both cases.

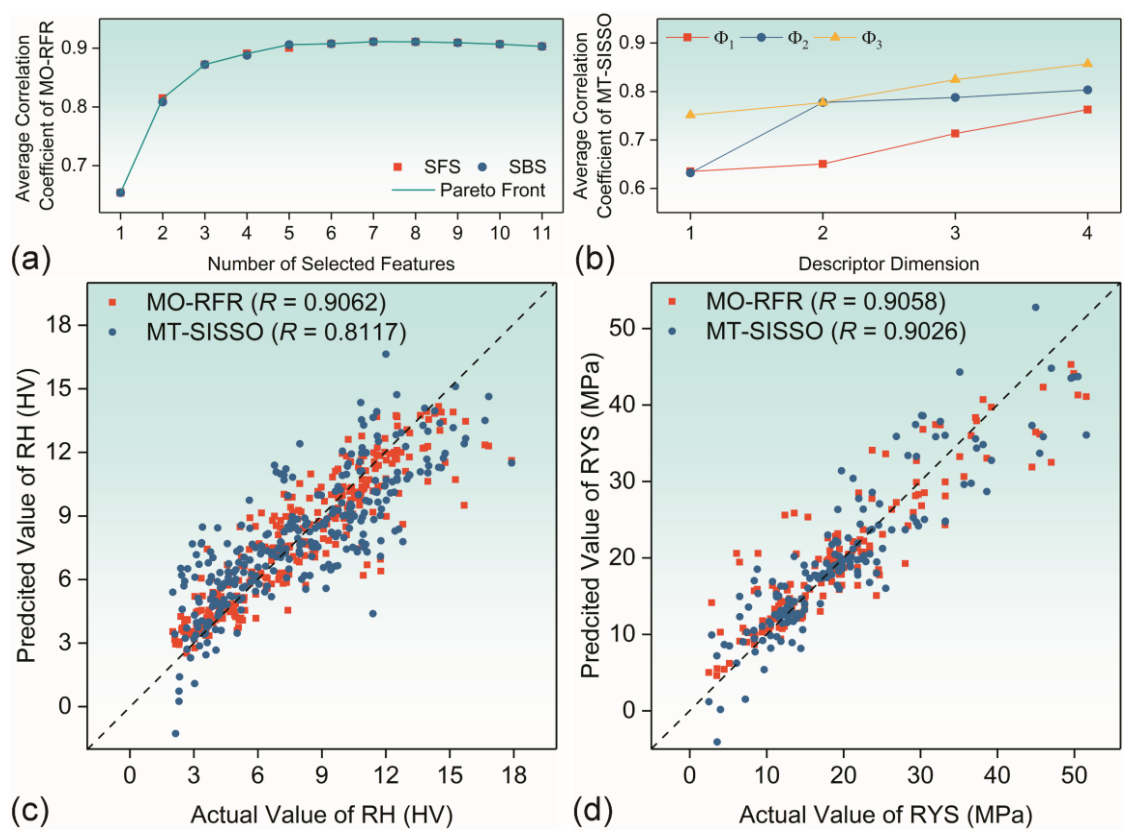

Figure 6. The process of feature selection and multi-objective random forest regression (MO-RFR) and multi-task SISSO (MT-SISSO) model construction for predicting both relative hardness (RH) and relative compressive yield stress (RYS) of CCAs. (a) shows the average correlation coefficient of MO-RFR models with 100 trees against a certain number of features selected by the sequential backward selection (SBS) and the sequential forward selection (SFS). (b) shows the average correlation coefficient of MT-SISSO models as a function of the dimension of the descriptor, for different size of the feature space. The (c) predicted RH and (d) predicted RYS values with MO-RFR (blue dots) and MT-SISSO (red dots) are plotted against the corresponding actual values.

The multi-task compressed-sensing method, MT-SISSO, proposed by Ouyang et. al.[37] is employed here to identify descriptors for *RH* and *RYS*. The average correlation coefficients of MT-SISSO models with different dimensions of descriptor, $\Omega$, for different sizes of feature space ($\Phi_1$, $\Phi_2$, and $\Phi_3$) are shown in Figure 6(b). Four descriptors in equations (12)-(15) constructed in the feature space $\Phi_3$ (~2.8 billion potential descriptor candidates) are found to affect both of *RH* and *RYS* as equation (16) and (17), respectively. The correlation coefficient of equations (16) and (17) can reach 0.8117 and 0.9026, respectively, shown as Figure 6(c) and 6(d).

$$\Omega_{41} = \frac{\sqrt[3]{H_f/\overline{AN}^2}}{\frac{\delta_{I1}}{\delta_{Tb}} + \frac{\delta_{I2}}{\delta_{MR}}} \tag{12}$$

$$\Omega_{42} = \frac{\delta_{Cp}^2}{\delta_{I1} - \delta_{Tb}} + \frac{\delta_{Tb}^2}{\delta_{Tb} - \delta_{MR}} \tag{13}$$

$$\Omega_{43} = \frac{(S_L/\overline{dVEC}) \cdot \exp(\overline{dVEC})}{\overline{dVEC}^3 \cdot \sqrt[3]{\delta_{I2}}} \tag{14}$$

$$\Omega_{44} = \frac{(\delta_{Cp} - \delta_{Tb}) \cdot \exp(\overline{dVEC})}{\ln(\overline{I2} \cdot \delta_{I2})} \tag{15}$$

$$RH = -3324.0962\Omega_{41} - 0.0394\Omega_{42} - 0.8448\Omega_{43} - 0.0002\Omega_{44} + 8.0622 \tag{16}$$

$$RYS = -8913.1997\Omega_{41} + 3.5791\Omega_{42} - 2.1305\Omega_{43} - 0.0065\Omega_{44} + 18.7985 \tag{17}$$

**Conclusion**

In this work, we proposed a ML framework to investigate the phase formation ability and predict the mechanical properties of CCAs. A dataset of 541 CCAs consisting of 61 AM samples, 133 SP-SS samples, 31 SP-IM samples, 92 MP-SS samples and 224 MP-IM samples was created by collecting data from related studies. Six features including $\overline{Tb}$, $\overline{dVEC}$, $\delta_{MR}$, $\delta_{Tb}$, $H_f$ and $S_L$ were found to affect the forming phases of CCAs as AM, SP, or MP. RF model built on these six features with 20 decision trees has an accuracy of 86.5 %. $\overline{Tb}$, $\overline{I2}$, $\delta_{I1}$, $\delta_{Cp}$, and $H_f$ determine the phase of crystalline CCAs as SS or IM, and the built RF models with 160 trees has an accuracy of 87.5 %. In addition, $\overline{AN}$, $\overline{dVEC}$, $\delta_{I2}$ and $\delta_{Cp}$ affect the structures in solid solutions of CCAs, and the built RF model with 260 trees has an accuracy of 90%. Moreover, $\overline{AN}$, $\overline{I2}$, $\overline{dVEC}$, $\delta_{I2}$ and $\delta_{MR}$ affect the hardness and compressive yield stress of CCAs, the correlation coefficients of predict both of these two properties are higher than 0.905. Three descriptors given by multi-task SISSO are found to influence relative hardness and relative compressive yield stress with a correlation coefficient of 0.8117 and 0.9026, respectively.

The present work demonstrates the great potential of machine learning approaches in the prediction of phase formations and mechanical properties of CCAs.

**Acknowledgments**

We would like to thank Doctor Ouyang Runhai for his expert advices on SISSO. This work was supported by the National Key Research and Development Program of China (No. 2018YFB0704400), the Hong Kong Polytechnic University (internal grant Nos. 1-ZE8R and G-YBDH), and the 111 Project (grant No. D16002) from the State Administration of Foreign Experts Affairs, PRC.